\documentclass[twocolumn]{jpsj3}
\usepackage{graphicx}
\usepackage{color}
\usepackage{bm}

\title{%
Gapless States Localized along a Staircase Edge
in Second-Order Topological Insulators
}

\author{%
Yuki Nagasato$^{1}$, Yositake Takane$^{1}$\thanks{E-mail: takane@hiroshima-u.ac.jp},
Yukinori Yoshimura$^{2}$, Shin Hayashi$^{2,3}$,\\
and Takeshi Nakanishi$^{2}$
}

\inst{%
$^1$Graduate School of Advanced Science and Engineering,\\
Hiroshima University, Higashihiroshima, Hiroshima 739-8530, Japan\\
$^2$AIST-TohokuU Mathematics for Advanced Materials Open Innovation Laboratory (MathAM-OIL), \\
National Institute of Advanced Industrial Science and Technology (AIST), \\
c/o Advanced Institute for Materials Research (AIMR), Tohoku University,\\ 
2-1-1 Katahira, Aoba, Sendai 980-8577, Japan\\
$^3$JST, PRESTO, 4-1-8 Honcho, Kawaguchi, Saitama 332-0012, Japan
}

\recdate{ \hspace{50mm} }

\abst{%
A second-order topological insulator on a two-dimensional square lattice hosts
zero-dimensional states inside a band gap.
They are localized near $90^{\circ}$ and $270^{\circ}$ corners
constituting an edge of the system.
When the edge is in a staircase form consisting of these two corners,
two families of edge states
(i.e., one-dimensional states localized near the edge)
appear as a result of the hybridization of zero-dimensional states.
We identify symmetry that makes them gapless.
We also show that a pair of nontrivial winding numbers
associated with this symmetry guarantee a gapless spectrum of edge states,
indicating that bulk--boundary correspondence holds
in this topological insulator with a staircase edge.
}

\begin{document}
\sloppy
\maketitle

\section{Introduction}

Recently, higher-order topological insulators have attracted
considerable attention.~\cite{benalcazar1,benalcazar2,langbehn,song,
hashimoto,schindler1,geier18,khalaf,ezawa1,ezawa2,fukui1,matsugatani,
hayashi1,hayashi2,trifunovic,araki,kunst19,wang19,fukui2,liu,okugawa,
takane,rui20,spurrier20,takahashi20,tanaka20PRB,roberts20,hashimoto20,
araki20,trifunovic20,claes20,ezawa20,asaga20,arai,po17,watanabe18,khalaf18X,
song18,tang19nature,zhang19,vergniory19,serra-garcia,peterson,
schindler2,imhof,noguchi21}
A two-dimensional second-order topological insulator hosts zero-dimensional
states at its corners, whereas a three-dimensional second-order
(third-order) topological insulator hosts one-dimensional
(zero-dimensional) states at its edges (corners).
That is, a $d$-dimensional $\mathcal{D}$th-order topological insulator
hosts $(d-\mathcal{D})$-dimensional states at its boundary,
where $2 \le \mathcal{D} \le d$.
These states appear as midgap states.

%%%%%%%%%%%%%%%%%%
\begin{figure}[bp]
\vspace{-5mm}
\hspace{-3mm}
\includegraphics[height=3.5cm]{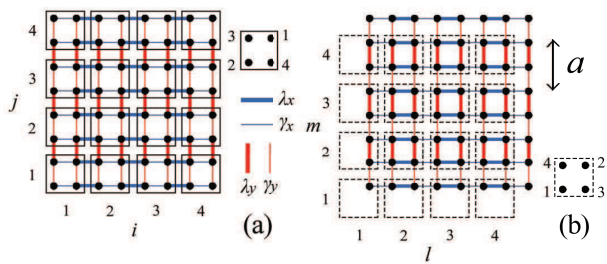}
\caption{
(Color online) Model system on a square lattice.
Thick horizontal (vertical) lines represent $\lambda_{x}$ ($\lambda_{y}$)
and thin horizontal (vertical) lines represent $\gamma_{x}$ ($\gamma_{y}$).
(a) Each solid square represents a unit cell, and
(b) each dotted square represents a dual cell.
}
\end{figure}
%%%%%%%%%%%%%%%%%%
A typical second-order topological insulator~\cite{benalcazar1,benalcazar2}
is defined on a two-dimensional square lattice,
in which each unit cell consists of four sites [see Fig.~1(a)].
This rectangular topological insulator
hosts zero-dimensional states at its corners.
For simplicity, the energy of a zero-dimensional state
in the large system-size limit is set equal to zero.
Let us focus on the system with a staircase edge as shown in Fig.~2.
Such an edge is characterized by a unit step of the staircase.
Let $N_{x}$ and $N_{y}$ be the numbers of unit cells
in each unit step in the $x$ and $y$ directions, respectively.
They are $(N_{x},N_{y}) = (1,1)$ in panel~(a),
$(2,1)$ in panel~(b), and $(3,2)$ in panel~(c).
We refer to a staircase edge with $(N_{x},N_{y})$
as a $(N_{x},N_{y})$ edge hereafter.
%%%%%%%%%%%%%%%%%%
\begin{figure}[bp]
\begin{tabular}{cc}
\begin{minipage}{0.4\hsize}
\begin{center}
\hspace{0mm}
\includegraphics[height=2.8cm]{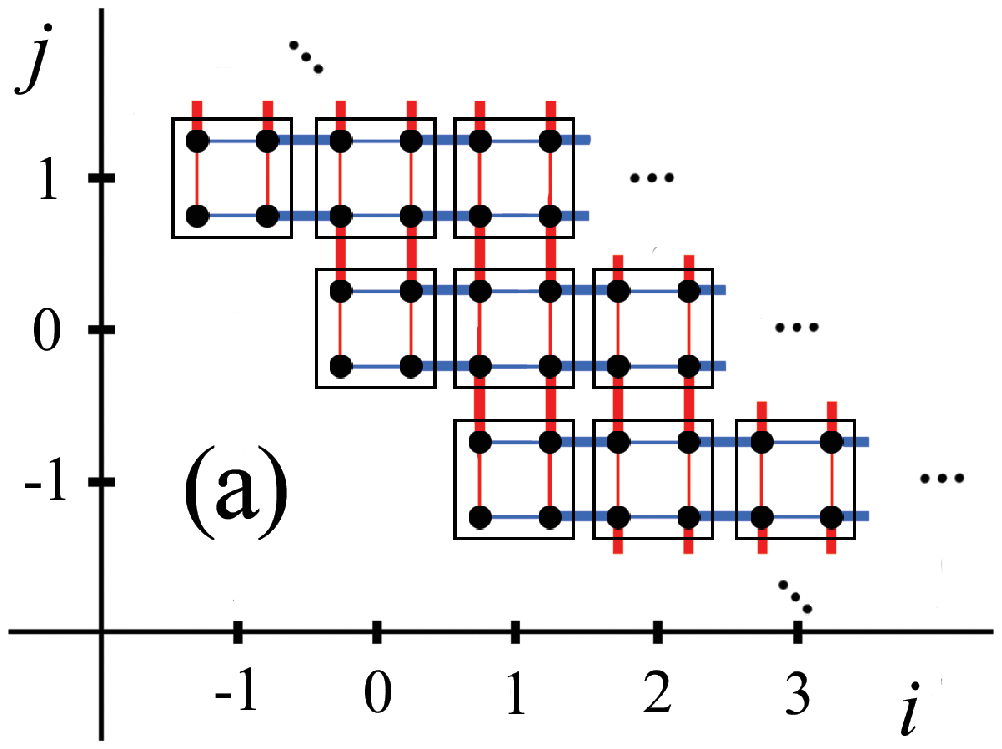}
\end{center}
\end{minipage}
\begin{minipage}{0.6\hsize}
\begin{center}
\hspace{0mm}
\includegraphics[height=2.8cm]{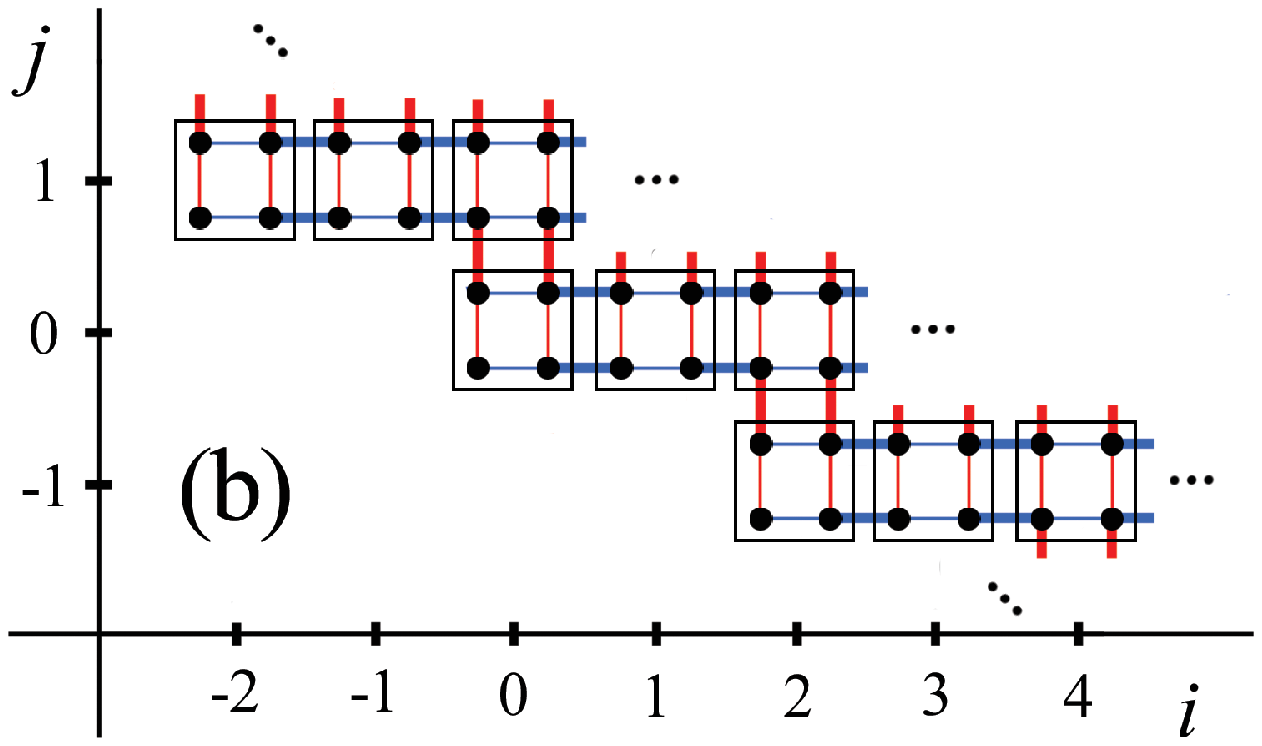}
\end{center}
\end{minipage}
\end{tabular}
\vspace{2mm}
\hspace{5mm}
\includegraphics[height=3.4cm]{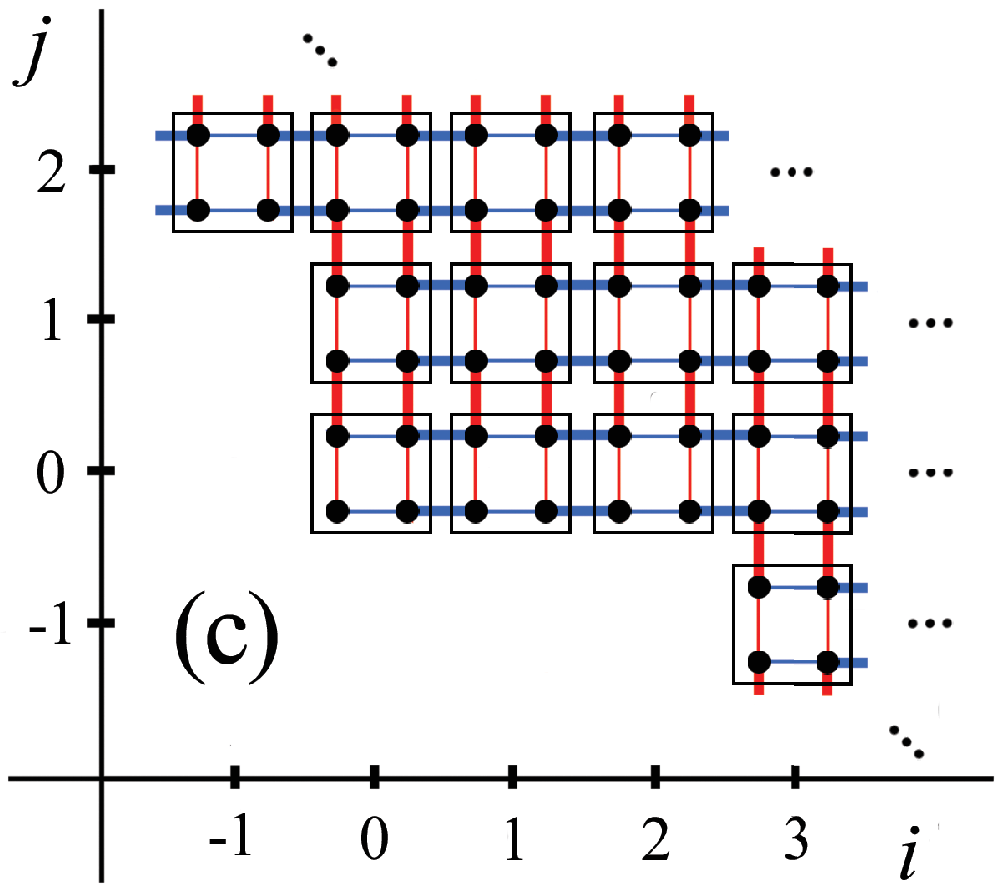}
\caption{
(Color online) Edge structures considered in the text.
(a) $(1,1)$ edge, (b) $(2,1)$ edge, and (c) $(3,2)$ edge.
Each unit cell represented by a solid square is characterized by
indices $i$ and $j$, respectively specifying its location
in the $x$ and $y$ directions.
}
\end{figure}
%%%%%%%%%%%%%%%%%%

As explained in the next section, in the system with a staircase edge,
two families of edge states (i.e., one-dimensional states localized
near the edge) appear: one originates from $90^{\circ}$ corners
whereas the other from $270^{\circ}$ corners.
A spectrum of the edge states tends to have a gap
owing to the hybridization between the two families.
However, numerical results (see Sect.~3)  show that the spectrum becomes
gapless under the condition given in Eq.~(\ref{eq:condition-ge}).
This suggests that a certain symmetry prohibits the hybridization
between the two families, leading to a gapless spectrum.
In this paper, we elucidate the reason why
the spectrum is gapless under that condition.
We identify the symmetry that prohibits the hybridization
between the two families and determine the wavefunctions of
a pair of zero-energy edge states at a gapless point.
We also show that the bulk--boundary correspondence holds in this system.
That is, the appearance of a pair of zero-energy edge states
is in a one-to-one correspondence with a pair of nontrivial winding numbers
associated with the symmetry.

This paper is organized as follows.
In the next section, we present a tight-binding Hamiltonian of
second-order topological insulators on a square lattice.
In Sect.~3, we numerically calculate the spectra of edge states
in ribbon-shaped systems with a pair of staircase edges
and give conditions for the appearance of gapless edge states.
In Sect.~4, we derive a one-dimensional model that is convenient
to study edge states localized along a $(1,1)$ edge
and identify the symmetry that protects a pair of zero-energy edge states.
The wavefunctions of two zero-energy edge states are analytically determined.
In Sect.~5, we show that the bulk--boundary correspondence
holds in this system.
In Sect.~6, the analysis given in the preceding two sections is extended
to the $(2,1)$ and $(3,2)$ edge cases.
The last section is devoted to summary and discussion.

\section{Model}

We present a tight-binding model for second-order topological insulators
on a square lattice with the lattice constant $a$, where the unit cell
consists of four sites numbered by $1$, $2$, $3$, and $4$
as shown in Fig.~1(a).
Each unit cell is characterized by indices $i$ and $j$ respectively
specifying its location in the $x$ and $y$ directions.
The four-component state vector for the $(i,j)$th unit cell is expressed as
\begin{align}
  |i,j \rangle
  =  \bigl\{ |i,j \rangle_{1}, |i,j \rangle_{2},
             |i,j \rangle_{3}, |i,j \rangle_{4} \bigr\} ,
\end{align}
where the subscript specifies the four sites.
The Hamiltonian is given by
$H = H_{\rm intra}+H_{\rm inter}$ with~\cite{benalcazar1,benalcazar2}
\begin{align}
   H_{\rm intra}
 & = \sum_{i,j} |i,j \rangle h_{\rm intra} \langle i,j| ,
         \\
   H_{\rm inter}
 & = \sum_{i,j} \bigl\{ |i+1,j \rangle h_{x}^{\lambda} \langle i,j|
                        + {\rm h.c.} \bigr\}
        \\
 & + \sum_{i,j} \bigl\{ |i,j+1 \rangle h_{y}^{\lambda} \langle i,j|
                        + {\rm h.c.} \bigr\} ,
\end{align}
where
\begin{align}
   h_{\rm intra}
 & = \left[ 
       \begin{array}{cccc}
         0 & 0 & \gamma_{x} & \gamma_{y} \\
         0 & 0 & -\gamma_{y} & \gamma_{x} \\
         \gamma_{x} & -\gamma_{y} & 0 & 0 \\
         \gamma_{y} & \gamma_{x} & 0 & 0 \\
       \end{array}
     \right] ,
               \\
   h_{x}^{\lambda}
 & = \left[ 
       \begin{array}{cccc}
         0 & 0 & 0 & 0 \\
         0 & 0 & 0 & \lambda_{x} \\
         \lambda_{x} & 0 & 0 & 0 \\
         0 & 0 & 0 & 0 \\
       \end{array}
     \right] ,
               \\
   h_{y}^{\lambda}
 & = \left[ 
       \begin{array}{cccc}
         0 & 0 & 0 & 0 \\
         0 & 0 & -\lambda_{y} & 0 \\
         0 & 0 & 0 & 0 \\
         \lambda_{y} & 0 & 0 & 0 \\
       \end{array}
     \right] .
\end{align}
The system described by the Hamiltonian is topologically nontrivial
under the condition~\cite{benalcazar1,benalcazar2}
\begin{align}
      \label{eq:TL}
    -\lambda_{x} < \gamma_{x} < \lambda_{x} ,
         \hspace{4mm}
    -\lambda_{y} < \gamma_{y} < \lambda_{y} ,
\end{align}
where $\lambda_{x}$ and $\lambda_{y}$ are assumed to be positive.
Without loss of generality,  we also assume that $\gamma_{x} \ge 0$
and $\gamma_{y} \ge 0$ (see the last paragraph of Sect.~3).
The model can be regarded as a two-dimensional extension of
the Su--Schrieffer--Heeger model.~\cite{su}
The Fourier transform of the Hamiltonian is
\begin{align}
      \label{eq:unit-cell-H}
   H(k_{x},k_{y})
 & = \lambda_{x}
     \left[ 
       \begin{array}{cc}
         0 & Q \\
         Q^{\dagger} & 0 \\
       \end{array}
     \right]
\end{align}
with
\begin{align}
   Q
  = \left[ 
       \begin{array}{cc}
         \frac{\gamma_{x}}{\lambda_{x}} + e^{ik_{x}a}
           & \alpha \left(\frac{\gamma_{y}}{\lambda_{y}} + e^{ik_{y}a}
                                                                  \right) \\
         -\alpha \left(\frac{\gamma_{y}}{\lambda_{y}} + e^{-ik_{y}a} \right)
           & \frac{\gamma_{x}}{\lambda_{x}} + e^{-ik_{x}a} \\
       \end{array}
     \right] ,
\end{align}
where
\begin{align}
   \alpha = \frac{\lambda_{y}}{\lambda_{x}} .
\end{align}

In addition to unit cells, we also use dual cells defined in Fig.~1(b)
to describe edge states.
The transfer of an electron between neighboring dual cells is
characterized by $\gamma_{x}$ and $\gamma_{y}$.
Therefore, each dual cell is disconnected from others
in the limit of $\gamma_{x}=\gamma_{y}=0$.
%%%%%%%%%%%%%%%%%%
\begin{figure}[btp]
\begin{center}
\includegraphics[height=8.0cm, angle=90]{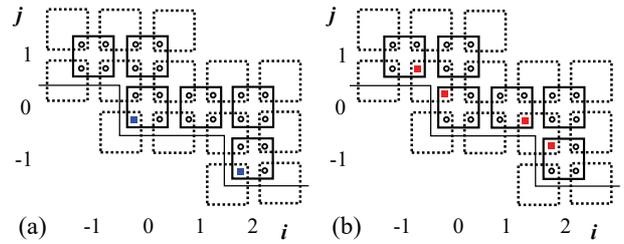}
\end{center}
\caption{
(Color online)
In the limit of $\gamma_{x}=\gamma_{y}=0$, a strongly localized
zero-energy state appears at each corner.
(a) A zero-energy state at a $90^{\circ}$ corner has a finite amplitude
only on the second site designated by filled squares (blue),
and (b) that at a $270^{\circ}$ corner has a finite amplitude
only on the third and fourth sites designated by filled squares (red).
}
\end{figure}
%%%%%%%%%%%%%%%%%%
As shown in Fig.~3, each dual cell located at a $90^{\circ}$ corner
contains only one site (i.e., a second site), indicating that such a site
is disconnected from other sites in the limit of $\gamma_{x}=\gamma_{y}=0$.
This means that, in the limit of $\gamma_{x}=\gamma_{y}=0$,
a strongly localized zero-energy state appears on the second site
in each dual cell located at a $90^{\circ}$ corner [see Fig.~3(a)].
These zero-energy states are combined to form one-dimensional edge states
along a staircase edge when $\gamma_{x}, \gamma_{y} \neq 0$.
Each dual cell located at a $270^{\circ}$ corner
contains three sites (i.e., second, third, and fourth sites), which are
disconnected from others in the limit of $\gamma_{x}=\gamma_{y}=0$.
It has been shown that, in the limit of $\gamma_{x}=\gamma_{y}=0$,
a strongly localized zero-energy state appears
on the third and fourth sites in each dual cell
located at a $270^{\circ}$ corner [see Fig.~3(b)].~\cite{takane}
These zero-energy states are also combined to form one-dimensional edge states
along a staircase edge when $\gamma_{x}, \gamma_{y} \neq 0$.
The above argument implies that two families of edge states
appear along a staircase edge as long as the system is
topologically nontrivial under the condition of Eq.~(\ref{eq:TL}).
However, this topological nontriviality does not ensure that
the edge states become gapless when $\gamma_{x}, \gamma_{y} \neq 0$.
A certain condition is necessary to make them gapless;
otherwise, they hybridize with each other
and an energy gap appears in the spectrum.

\section{Numerical Results}

To elucidate conditions for the appearance of gapless edge states,
we examine the spectra of ribbon-shaped systems with a pair of staircase edges
as shown in Fig.~4.
In this figure, a dashed rectangle designates a unit cell of
the ribbon system consisting of $N$ unit squares in the horizontal direction,
and $e^{i\theta}$ represents a phase difference of an eigenfunction
between two neighboring unit cells.
%%%%%%%%%%%%%%%%%%
\begin{figure}[btp]
\begin{center}
\includegraphics[height=2.0cm]{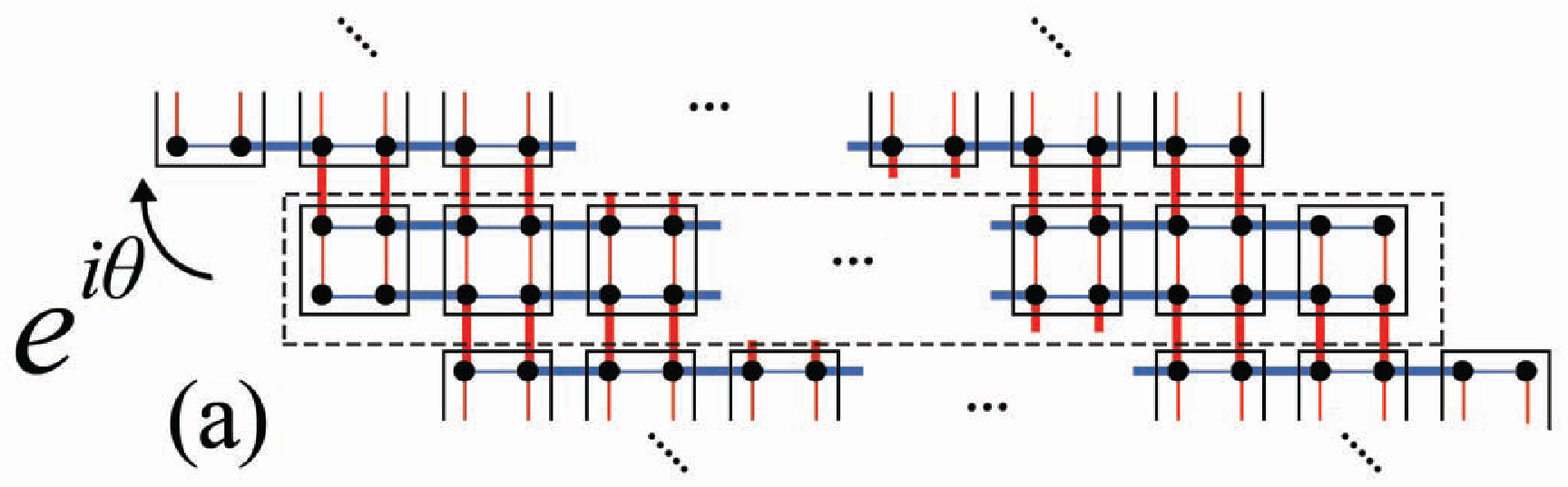}
\vspace{2mm}
\includegraphics[height=2.0cm]{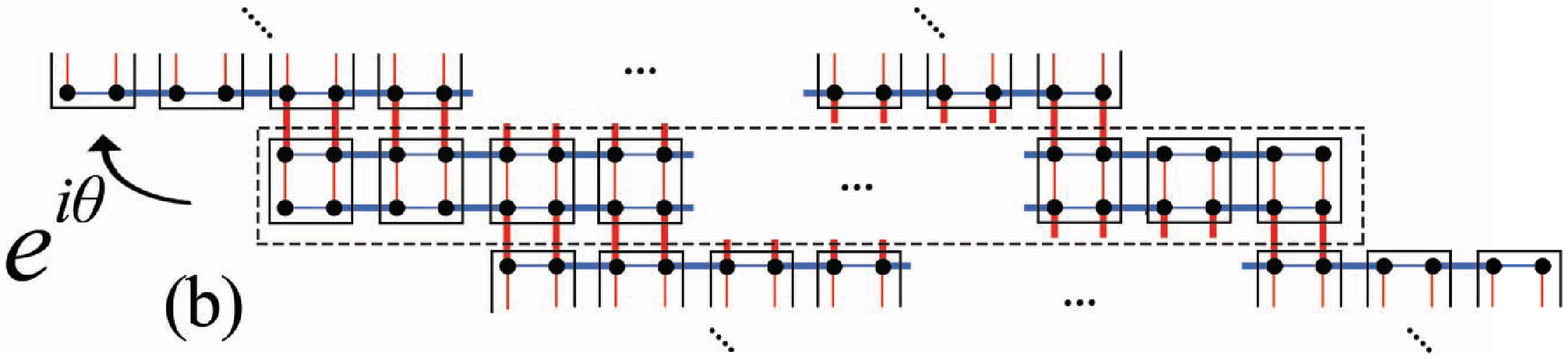}
\end{center}
\caption{(Color online)
Ribbon systems with (a) $(1,1)$ edges and (b) $(2,1)$ edges.
}
\end{figure}
%%%%%%%%%%%%%%%%%%
We set $\alpha = 1$ throughout this section.
Spectra in two ribbon systems with $N = 10$ are shown in Fig.~5.
Panel~(a) shows the spectrum in the $(1,1)$ edge case
with $\gamma_{x}/\lambda_{x} = \gamma_{y}/\lambda_{y} = 0.5$,
in which midgap edge states have a gapless point at $\theta = 0$.
Panel~(b) shows the spectrum in the $(2,1)$ edge case with
$(\gamma_{x}/\lambda_{x})^{2} = \gamma_{y}/\lambda_{y} = 0.25$,
in which midgap edge states have a gapless point at $\theta = \pi$.
In both cases,
four zero-energy edge states appear at the gapless point
because the two branches near $E = 0$ are doubly degenerate.
In the $(2,1)$ edge case, the probability density distributions of
two zero-energy edge states near the left edge of the ribbon system
are shown in Fig.~6, where the probability density distribution
in panel~(a) is localized near $90^{\circ}$ corners
and that in panel~(b) is localized near $270^{\circ}$ corners.
Let us consider a gap $E_{\rm G}$ of the edge states
in a space of $\gamma_{x}/\lambda_{x}$ and $\gamma_{y}/\lambda_{y}$.
Here, $E_{\rm G}$ is defined as the minimum value of vertical separation
between the upper and lower branches of the edge states
as a function of $\theta$.
Color plots of $E_{\rm G}$ normalized by $\lambda_{x}$
are shown in Fig.~7 for two ribbon systems.
Panel~(a) shows $E_{\rm G}$ in the $(1,1)$ edge case with $N = 200$,
which vanishes on the dashed line representing
$\gamma_{x}/\lambda_{x} = \gamma_{y}/\lambda_{y}$.
Panel~(b) shows $E_{\rm G}$ in the $(2,1)$ edge case with $N = 300$,
which vanishes on the dashed line representing
$(\gamma_{x}/\lambda_{x})^{2} = \gamma_{y}/\lambda_{y}$.

The numerical results shown above
suggest that the edge states become gapless when
\begin{align}
   \label{eq:condition-ge}
 \left(\frac{\gamma_{x}}{\lambda_{x}}\right)^{N_{x}}
 = \left(\frac{\gamma_{y}}{\lambda_{y}}\right)^{N_{y}} < 1
\end{align}
at $\theta = 0$ ($\theta = \pi$) if $N_{x}+N_{y}$ is even (odd).
This is also supported by numerical results for the ribbon system
with $(3,2)$ edges (data not shown).
The presence of a gapless point at $\theta = 0$ ($\theta = \pi$) indicates
that a pair of zero-energy edge states are invariant (change their sign)
under the lattice translation of $(\pm N_{x}a,\mp N_{y}a)$.

If $\gamma_{x}$ and $\gamma_{y}$ are allowed to be negative,
a gapless point appears when
$|\gamma_{x}/\lambda_{x}|^{N_{x}} = |\gamma_{y}/\lambda_{y}|^{N_{y}} < 1$
at $\theta = 0$ ($\theta = \pi$) if $\kappa >0$ ($\kappa <0$), where
$\kappa = (-\gamma_{x}/\lambda_{x})^{N_{x}}(-\gamma_{y}/\lambda_{y})^{N_{y}}$.
This indicates that the allowance of $\gamma_{x} < 0$ and/or $\gamma_{y} < 0$
does not lead to a qualitative change in the behavior of edge states.
%%%%%%%%%%%%%%%%%%
\begin{figure}[btp]
\begin{center}
\includegraphics[height=3.6cm]{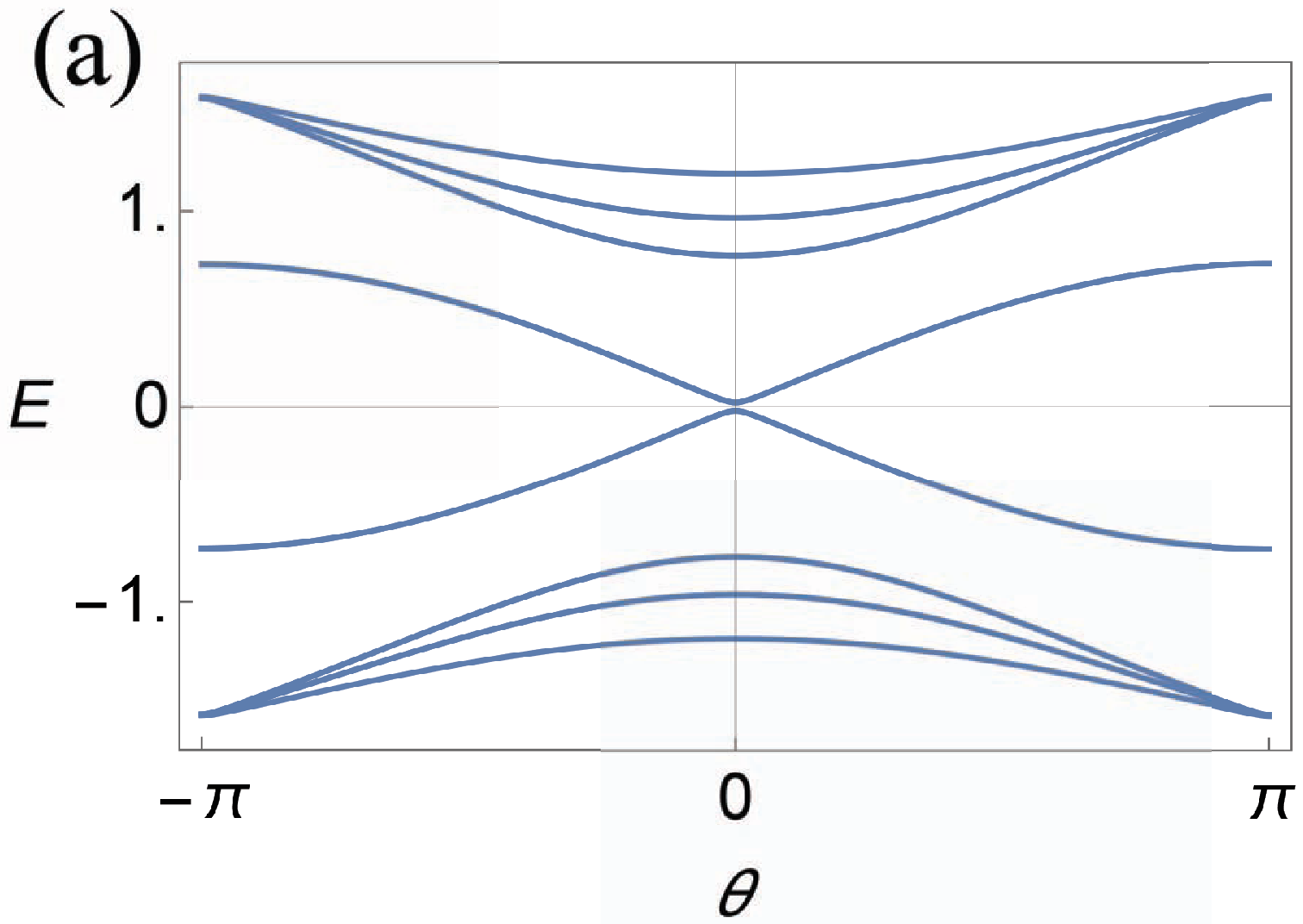}
\includegraphics[height=3.6cm]{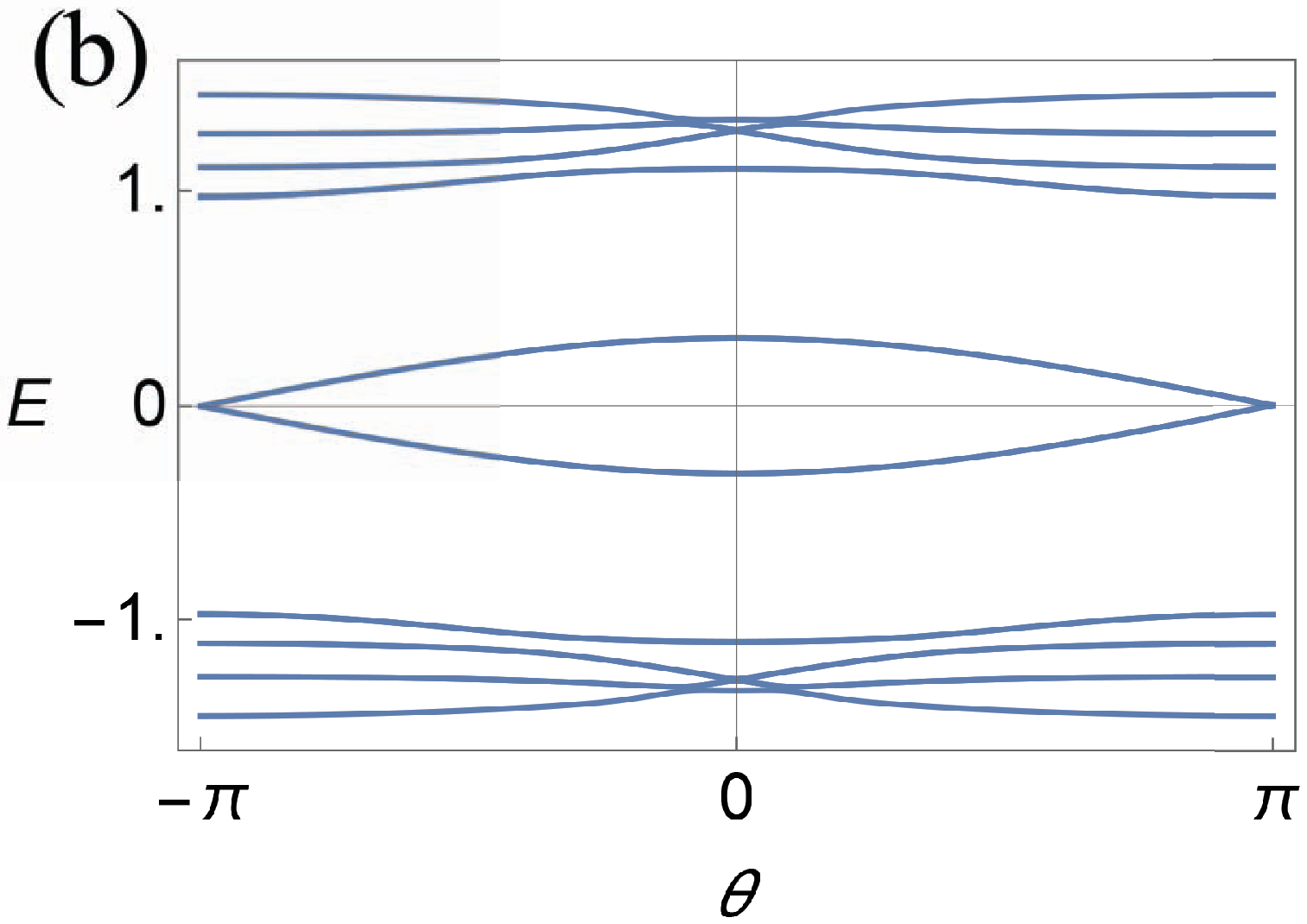}
\end{center}
\caption{(Color online)
Spectra of the ribbon system in the (a) $(1,1)$ edge case with
$\gamma_{x}/\lambda_{x} = \gamma_{y}/\lambda_{y} = 0.5$
and (b) $(2,1)$ edge case with
$(\gamma_{x}/\lambda_{x})^{2} = \gamma_{y}/\lambda_{y} = 0.25$.
}
\end{figure}
%%%%%%%%%%%%%%%%%%
%%%%%%%%%%%%%%%%%%
\begin{figure}[btp]
\begin{tabular}{cc}
\begin{minipage}{0.5\hsize}
\begin{center}
\includegraphics[height=2.1cm]{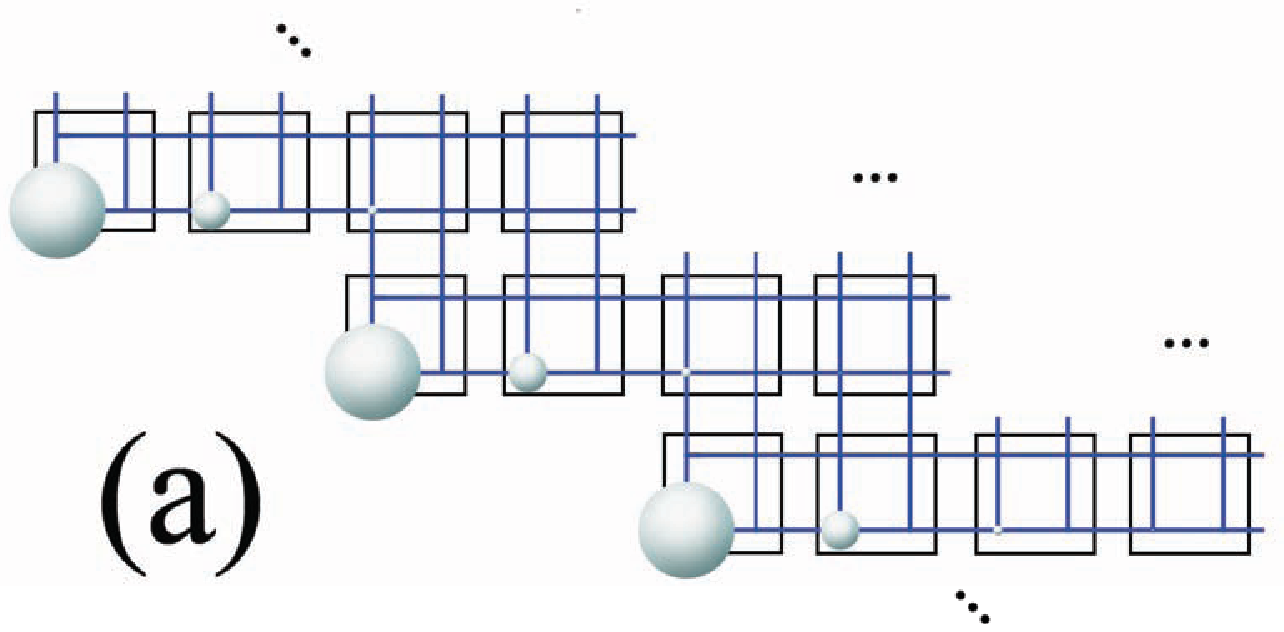}
\end{center}
\end{minipage}
\begin{minipage}{0.5\hsize}
\begin{center}
\includegraphics[height=2.1cm]{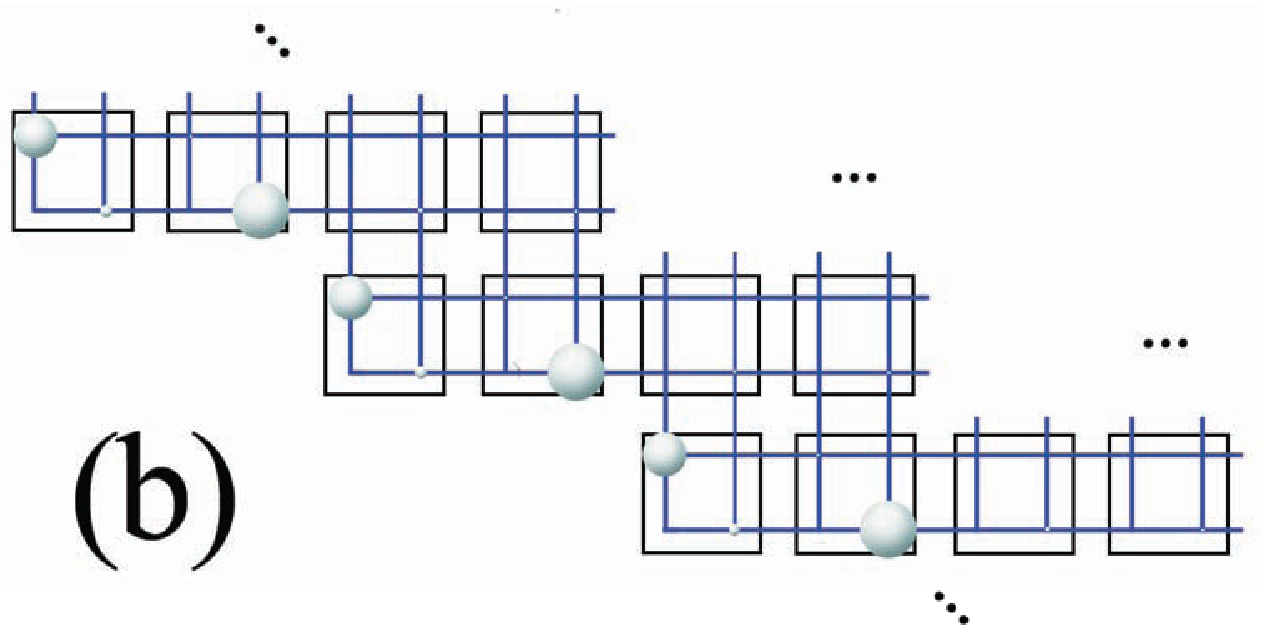}
\end{center}
\end{minipage}
\end{tabular}
\caption{(Color online)
Probability density distributions of two zero-energy edge states
in the $(2,1)$ edge case
with $(\gamma_{x}/\lambda_{x})^{2} = \gamma_{y}/\lambda_{y} = 0.25$,
where a probability density at each site is represented by the radius
of a sphere on the site.
The probability density distribution in (a)
is localized near $90^{\circ}$ corners,
whereas that in (b) is localized near $270^{\circ}$ corners.
}
\end{figure}
%%%%%%%%%%%%%%%%%%
%%%%%%%%%%%%%%%%%%
\begin{figure}[btp]
\begin{center}
(a)
\includegraphics[height=4.8cm]{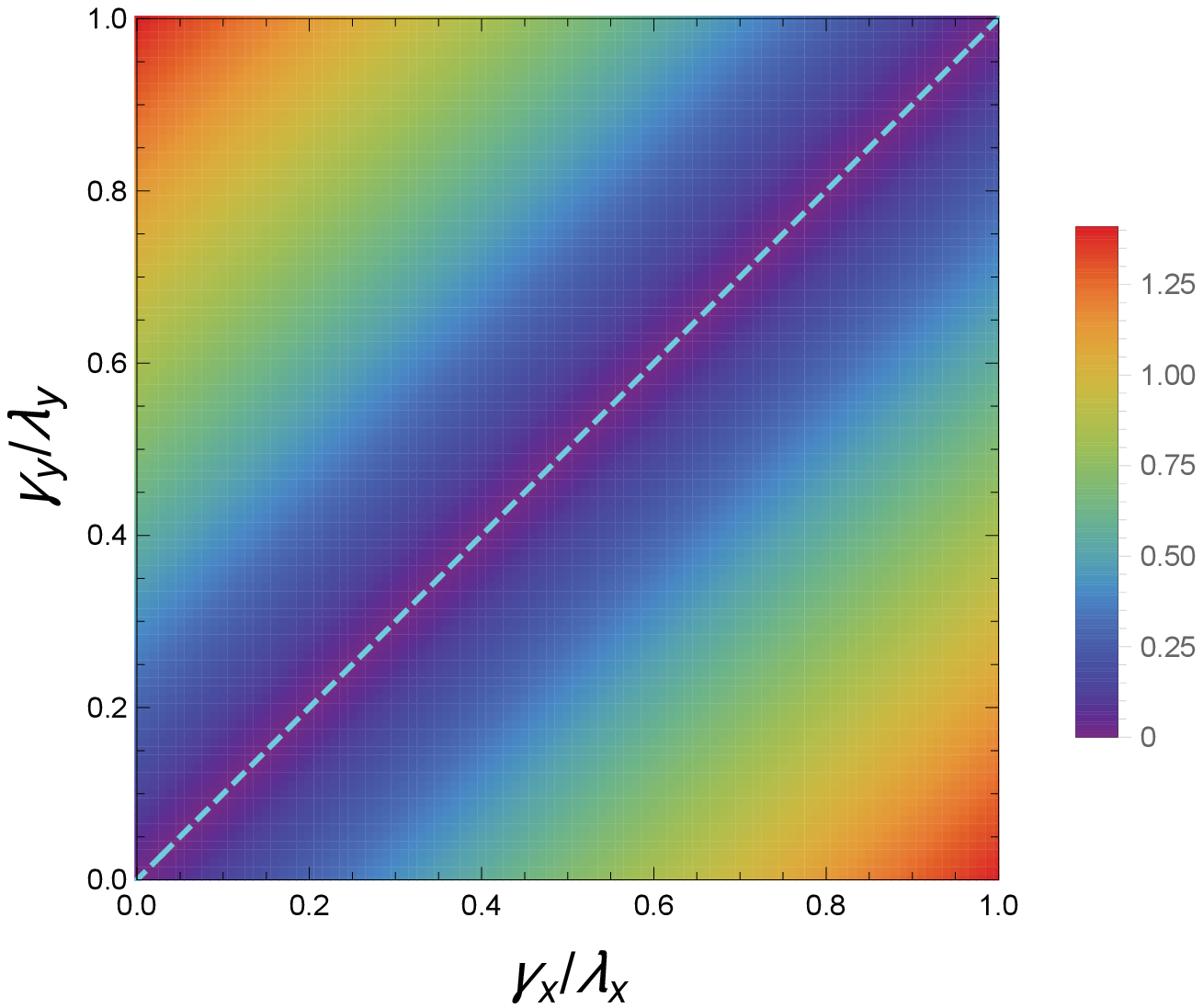}\\
(b)
\includegraphics[height=4.8cm]{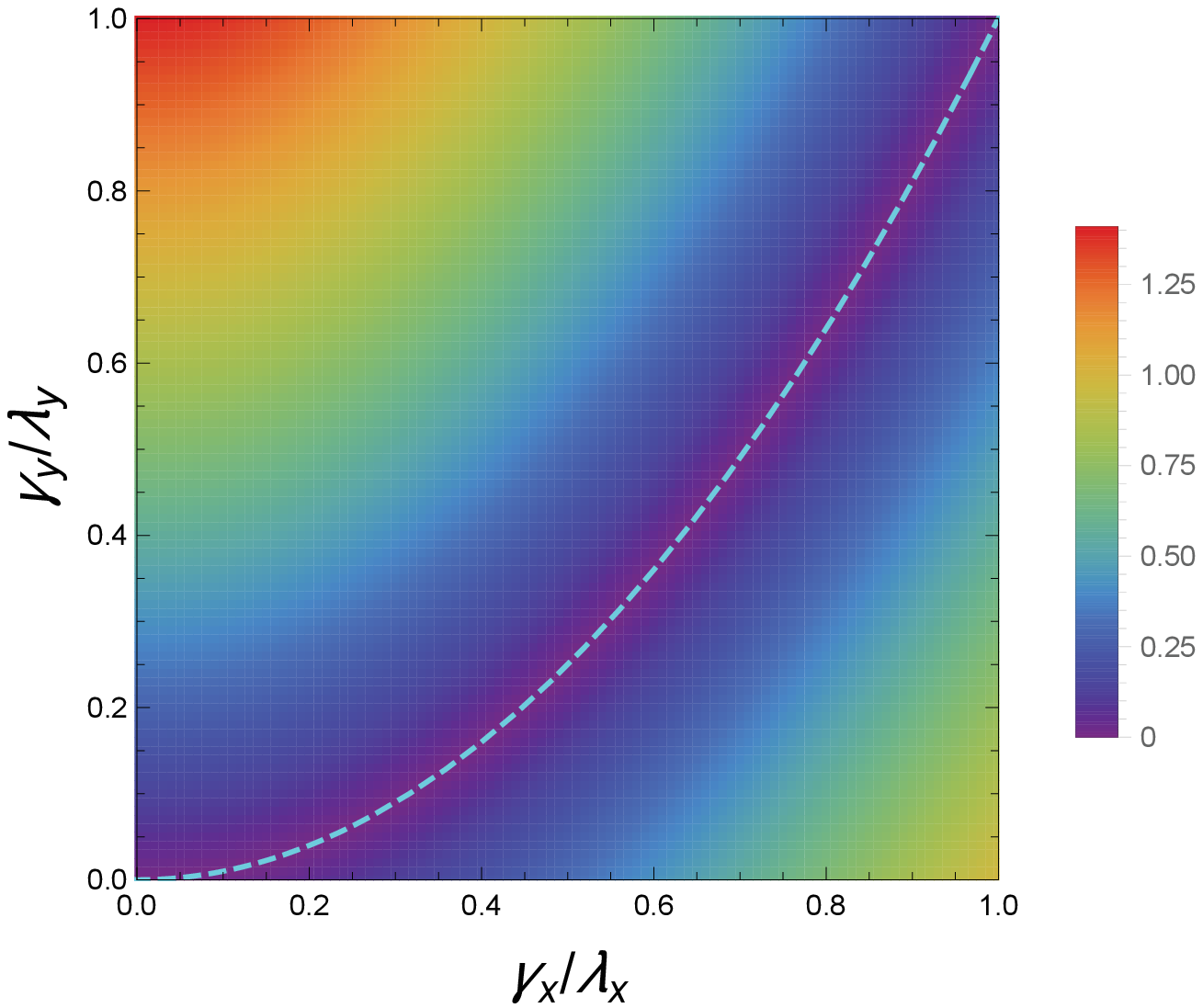}
\end{center}
\caption{(Color)
Gap between upper and lower branches of the edge states
in the space of $\gamma_{x}/\lambda_{x}$ and $\gamma_{y}/\lambda_{y}$
in the (a) $(1,1)$ edge case with $N = 200$
and (b) $(2,1)$ edge case with $N = 300$.
}
\end{figure}
%%%%%%%%%%%%%%%%%%

\section{Formulation for the Simplest Case}

Focusing on the system with a $(1,1)$ edge as shown in Fig.~2(a),
we show that the edge states become gapless
under the condition suggested in Sect.~3.
We present a theoretical framework to identify the symmetry
that makes them gapless.

The numerical results in Sect.~3 suggest that the edge states
become gapless under the condition
\begin{eqnarray}
   \label{eq:condition1}
 \frac{\gamma_{x}}{\lambda_{x}} = \frac{\gamma_{y}}{\lambda_{y}}= r
\end{eqnarray}
with $0 \le r < 1$,
and that two edge states at zero energy are invariant
under the lattice translation of $(\pm a,\mp a)$.
To describe the two zero-energy edge states localized along the $(1,1)$ edge,
we impose Eq.~(\ref{eq:condition1}) on $H(k_{x},k_{y})$
and replace both $k_{x}$ and $k_{y}$ with $k$ to take the invariance
under the lattice translation into account.
This results in
\begin{align}
   Q
  = \left[ 
       \begin{array}{cc}
         r + e^{ika} & \alpha \left(r+ e^{ika} \right) \\
         -\alpha \left(r+ e^{-ika} \right) & r + e^{-ika} \\
       \end{array}
     \right] .
\end{align}
It is convenient to rewrite it as
\begin{align}
      \label{eq:def-Q1}
   Q
  = \left[ 
       \begin{array}{cc}
         \Sigma \eta & \Gamma \eta \\
         -\Gamma^{*}\eta^{*} & \Sigma^{*}\eta^{*} \\
       \end{array}
     \right] ,
\end{align}
where~\cite{comment1}
\begin{align}
  \eta & = r + e^{ika} , \\
  \Sigma & = 1 , \\
  \Gamma & = \alpha .
\end{align}
Here, we define $H(k)$ as
\begin{align}
  H(k) = \left. H(k_{x},k_{y}) \right.|_{k_{x}=k,k_{y}=k},
\end{align}
which is referred to as a one-dimensional model
since it contains only one wavenumber $k$.
At $\alpha = 1$, it is equivalent to the model
introduced in Refs.~\citen{liu} and \citen{imhof}.
However, one-dimensional edge states are beyond the scope of these studies.

For this one-dimensional Hamiltonian,
let us introduce the following $4 \times 4$ unitary matrix:
\begin{align}
 U
  = \left[ 
       \begin{array}{cc}
         \sigma_{z} & 0 \\
         0 & \Lambda \\
       \end{array}
     \right]
\end{align}
with $\sigma_{z}$ being the $z$-component of Pauli matrices and
\begin{align}
 \Lambda
   = \frac{1}{\Theta}
   \left[ 
       \begin{array}{cc}
         |\Sigma|^{2}-|\Gamma|^{2} & 2\Sigma^{*} \Gamma \\
          2\Sigma \Gamma^{*} &  -|\Sigma|^{2}+|\Gamma|^{2} \\
       \end{array}
     \right] ,
\end{align}
where
\begin{align}
 \Theta = |\Sigma|^{2}+|\Gamma|^{2} .
\end{align}
By using $\sigma_{z}Q\Lambda = Q$, we can easily show that
$H(k)$ is invariant
under the unitary transformation described by $U$:
\begin{align}
U^{\dagger}H(k)U = H(k) .
\end{align}
In the limit of $\alpha = 1$,
$U$ represents the mirror-rotation symmetry.~\cite{imhof}

Because $U^{2} = 1_{4 \times 4}$, the eigenvalues of $U$ are $\pm 1$.
Two of the eigenvectors of $U$
corresponding to the eigenvalue of $1$ are given by
\begin{align}
 v_{+}^{(1)} = \left( \begin{array}{c}
                        1 \\ 0 \\ 0\\ 0 
                      \end{array} \right) ,
     \hspace{7mm}
 v_{+}^{(2)} = \frac{1}{\sqrt{\Theta}}
               \left( \begin{array}{c}
                        0 \\ 0 \\ \Sigma^{*} \\ \Gamma^{*}
                      \end{array} \right) ,
\end{align}
whereas the other two corresponding to the eigenvalue of $-1$ are
\begin{align}
 v_{-}^{(1)} = \left( \begin{array}{c}
                        0 \\ 1 \\ 0\\ 0 
                      \end{array} \right) ,
     \hspace{7mm}
 v_{-}^{(2)} = \frac{1}{\sqrt{\Theta}}
               \left( \begin{array}{c}
                        0 \\ 0 \\ \Gamma \\ -\Sigma
                      \end{array} \right) .
\end{align}
By using the unitary matrix given by
\begin{align}
 V = \left[ v_{+}^{(1)}, v_{+}^{(2)}, v_{-}^{(1)}, v_{-}^{(2)} \right] ,
\end{align}
we can transform the Hamiltonian as
\begin{align}
  V^{\dagger} H(k) V
  = \left[ \begin{array}{cc}
             H_{+} & 0 \\
             0 & H_{-}
           \end{array}
    \right]
\end{align}
with
\begin{align}
     \label{eq:H+}
  H_{+} & = \lambda_{x} \sqrt{\Theta}
            \left[ \begin{array}{cc}
                     0 & \eta \\
                     \eta^{*} & 0
                    \end{array}
            \right]  , \\
     \label{eq:H-}
  H_{-} & = \lambda_{x} \sqrt{\Theta}
            \left[ \begin{array}{cc}
                     0 & -\eta^{*} \\
                     -\eta & 0
                   \end{array}
            \right] .
\end{align}
Here, $H_{\pm}$ describes eigenstates in the subspace
spanned by $v_{\pm}^{(1)}$ and $v_{\pm}^{(2)}$.
As noted in Sect.~5, $H_{\pm}$ possesses a chiral symmetry.

Now, we show that two zero-energy edge states localized along the $(1,1)$ edge
are obtained by solving $H_{+}|\psi_{+}\rangle = 0$
and $H_{-}|\psi_{-}\rangle = 0$ after the replacement of
$e^{ika}$ with $\rho$ and $e^{-ika}$ with $\rho^{-1}$.
Because $\rho$ characterizes the attenuation of a solution
away from the edge, it must satisfy $|\rho| < 1$.
From $H_{-}|\psi_{-}\rangle = 0$, we find a solution with $\rho = -r$ as
\begin{align}
     \label{eq:psi_{-}}
 |\psi_{-}\rangle
    = \sum_{i,j} (-r)^{i+j} \, |i,j\rangle
      \left( \begin{array}{c}
               1 \\ 0
             \end{array} \right) ,
\end{align}
which does not amplify or attenuate along the $(1,1)$ edge.
Replacing $^{t}(1, 0)$ with the corresponding eigenvector $v_{-}^{(1)}$
in the original $4 \times 4$ space, we rewrite Eq.~(\ref{eq:psi_{-}}) as
\begin{align}
      \label{eq:psi_{-}(1,1)}
 |\psi_{-}\rangle
    = \sum_{i,j} (-r)^{i+j} \, |i,j\rangle
      \left( \begin{array}{c}
               0 \\ 1 \\ 0\\ 0
             \end{array} \right) .
\end{align}
It is easy to show that $|\psi_{-}\rangle$ satisfies
the boundary condition at the $(1,1)$ edge, that is, it obeys
the original eigenvalue equation of $H|\psi_{-}\rangle = 0$
on every site on the $(1,1)$ edge.
We observe that $|\psi_{-}\rangle$ vanishes in the limit of $r \to 0$
except on the second site in each dual cell located at a $90^\circ$ corner.
This clearly indicates that $|\psi_{-}\rangle$
originates from $90^\circ$ corners.

From $H_{+}|\psi_{+}\rangle = 0$, we find a solution with $\rho = -r$ as
\begin{align}
      \label{eq:psi_{+}}
 |\psi_{+}\rangle
    = \sum_{i,j} (-r)^{i+j} \, |i,j\rangle
      \left( \begin{array}{c}
               0 \\ 1
             \end{array} \right) .
\end{align}
Replacing $^{t}(0, 1)$ with the corresponding eigenvector $v_{+}^{(2)}$
in the original $4 \times 4$ space, we rewrite Eq.~(\ref{eq:psi_{+}}) as
\begin{align}
      \label{eq:psi_{+}(1,1)}
 |\psi_{+}\rangle
    = \sum_{i,j} (-r)^{i+j} \, |i,j \rangle
      \left( \begin{array}{c}
               0 \\ 0 \\ 1\\ \alpha
             \end{array} \right) ,
\end{align}
which also satisfies the boundary condition; it satisfies
$H|\psi_{+}\rangle = 0$ on every site on the $(1,1)$ edge.
We observe that $|\psi_{+}\rangle$ vanishes in the limit of $r \to 0$
except on the third and fourth sites
in each dual cell located at a $270^\circ$ corner.
This clearly indicates that $|\psi_{+}\rangle$
originates from $270^\circ$ corners.

\section{Bulk--Boundary Correspondence}

We show that the appearance of a zero-energy edge state described by $H_{\pm}$
is governed by a winding number associated with $H_{\pm}$, relying on
the argument of Ryu and Hatsugai~\cite{ryu} based on a chiral symmetry.
Note that $H_{\pm}$ possesses the chiral symmetry
\begin{align}
  \sigma_{z} H_{\pm} \sigma_{z} = - H_{\pm} .
\end{align}
This symmetry ensures that except at zero energy, edge states
appear in pairs: if one state has energy $\epsilon$,
the other has energy $-\epsilon$.
Conversely, it ensures that an unpaired edge state
must be at zero energy.

Let us consider the $(1,1)$ edge case with focus on
the zero-energy edge state of Eq.~(\ref{eq:psi_{-}(1,1)})
described by $H_{-}$ with $\eta = r + e^{ika}$ and $\Theta = 1 + \alpha^{2}$.
The winding number associated with $H_{-}$ is defined by
\begin{align}
      \label{eq:def_w-}
  w_{-} = \frac{1}{2\pi}\bigl.{\rm arg}\bigl\{-\sqrt{\Theta}\eta(k)^{*}\bigr\}
                        \bigr|^{\frac{2\pi}{a}}_{k=0} ,
\end{align}
which counts how many times the trajectory of $-\sqrt{\Theta}\eta(k)^{*}$
winds around the origin in the anti-clockwise direction
when $k$ varies from $0$ to $\frac{2\pi}{a}$.
It is given by
\begin{align}
  w_{-} & =  \left\{ \begin{array}{cc}
                         -1, & 0 \le r < 1 ,\\
                         0, & 1 < r .
                     \end{array}
             \right.
\end{align}
Let us consider the special case of $r = 0$ (i.e., $\gamma_{x}=\gamma_{y}=0$)
at which $w_{-} = -1$.
In this case, strongly localized zero-energy states appear
at all $90^{\circ}$ corners of the $(1,1)$ edge,
and they can form a zero-energy edge state
that is invariant under the lattice translation of $(\pm a,\mp a)$.
This state is equivalent to Eq.~(\ref{eq:psi_{-}(1,1)})
in the limit of $r \to 0$,
whereas its counterpart originating from $270^{\circ}$ corners
is equivalent to Eq.~(\ref{eq:psi_{+}(1,1)}) in the limit of $r \to 0$.
This means that the zero-energy state at $r = 0$
is unpaired in the subspace governed by $H_{-}$.
If $r$ is varied from $0$ under the condition of $0 \le r < 1$
and thus, $w_{-} = -1$ and an energy gap does not close,
its energy must stay at $E = 0$ as a result of the chiral symmetry.
We conclude that one edge state originating from $90^{\circ}$ corners
appears at zero energy as long as $w_{-} = -1$.

The above argument is also applicable to the zero-energy edge state
given in Eq.~(\ref{eq:psi_{+}(1,1)}) described by $H_{+}$.
The corresponding winding number $w_{+}$ is defined by
\begin{align}
       \label{eq:def_w+}
  w_{+} = \frac{1}{2\pi}\bigl.{\rm arg}\bigl\{\sqrt{\Theta}\eta(k)\bigr\}
                        \bigr|^{\frac{2\pi}{a}}_{k=0} ,
\end{align}
which results in
\begin{align}
  w_{+} =  \left\{ \begin{array}{cc}
                       1, & 0 \le r < 1 ,\\
                       0, & 1 < r .
                   \end{array}
           \right.
\end{align}
The one edge state originating from $270^{\circ}$ corners
appears at zero energy as long as $w_{+} = 1$.

The above analysis clearly indicates that the two zero-energy edge states
are protected by the symmetry represented by $U$
under the condition of Eq.~(\ref{eq:condition1}).
For arbitrary $\alpha$,
we refer to this as a generalized mirror-rotation symmetry.
Once this symmetry is broken,
the two zero-energy edge states hybridize with each other
and the energy of resulting states should deviate from zero.

\section{Application to Other Cases}

The analysis in the preceding two sections is straightforwardly
extended to an arbitrary $(N_{x},N_{y})$ edge.
In this section,
we show this taking the $(2,1)$ and $(3,2)$ edge cases as examples.

Let us consider the $(2,1)$ edge as shown in Fig.~2(b).
The numerical results in Sect.~3 show that the edge states
become gapless under the condition
\begin{align}
   \label{eq:condition2}
 \left(\frac{\gamma_{x}}{\lambda_{x}}\right)^{2}
 = \frac{\gamma_{y}}{\lambda_{y}}= r^{2}
\end{align}
with $0 \le r < 1$, and that two edge states at zero energy change
their sign under the lattice translation of $(\pm 2a,\mp a)$.
To describe the zero-energy edge states,
we impose Eq.~(\ref{eq:condition2}) on $H(k_{x},k_{y})$.
Moreover, we replace $k_{x}$ with $k$ and $k_{y}$ with $2k+k_{\pi}$
with $k_{\pi} = \frac{\pi}{a}$
such that the resulting Hamiltonian describes an eigenfunction
that changes its sign under the lattice translation of $(\pm 2a,\mp a)$
without changing its amplitude.
This results in
\begin{align}
      \label{eq:Q2-init}
   Q
  = \left[ 
       \begin{array}{cc}
         r + e^{ika} & \alpha \left(r^{2} - e^{i2ka} \right) \\
         -\alpha \left(r^{2} - e^{-i2ka} \right) & r + e^{-ika} \\
       \end{array}
     \right] .
\end{align}
It is convenient to rewrite it as
\begin{align}
   Q
  = \left[ 
       \begin{array}{cc}
         \Sigma \eta & \Gamma \eta \\
         -\Gamma^{*}\eta^{*} & \Sigma^{*}\eta^{*} \\
       \end{array}
     \right] ,
\end{align}
where
\begin{align}
  \eta  = & r + e^{ika} , \\
  \Sigma = & 1 , \\
  \Gamma = & \alpha \left(r - e^{ika}\right) .
\end{align}

The one-dimensional Hamiltonians $H_{+}$ and $H_{-}$ given in
Eqs.~(\ref{eq:H+}) and (\ref{eq:H-}), respectively, 
are also applicable to this case.
Again, we show that two zero-energy edge states are obtained by solving
$H_{+}|\psi_{+}\rangle = 0$ and $H_{-}|\psi_{-}\rangle = 0$
after the replacement of $e^{ika}$ with $\rho$ and $e^{-ika}$ with $\rho^{-1}$.
From $H_{-}|\psi_{-}\rangle = 0$, we find a solution with $\rho = -r$
and the eigenvector $^{t}(1, 0)$ corresponding to $v_{-}^{(1)}$.
This is expressed as
\begin{align}
      \label{eq:psi_{-}(2,1)}
 |\psi_{-}\rangle
    = \sum_{i,j} (-r)^{i}(-r^{2})^{j} \, |i,j \rangle
      \left( \begin{array}{c}
               0 \\ 1 \\ 0\\ 0
             \end{array} \right) ,
\end{align}
which satisfies the boundary condition at the $(2,1)$ edge.
We observe that $|\psi_{-}\rangle$ vanishes in the limit of $r \to 0$
except on the second site in each dual cell located at a $90^\circ$ corner.

To find another zero-energy edge state originating from $270^\circ$ corners,
we solve $H_{+}|\psi_{+}\rangle = 0$.
A solution with $\rho = -r$ and the eigenvector $^{t}(0, 1)$
corresponding to $v_{+}^{(2)}$ is expressed as
\begin{eqnarray}
 |\psi_{+}\rangle
    = \sum_{i,j} (-r)^{i}\left(-r^{2}\right)^{j} \, |i,j \rangle
      \left( \begin{array}{c}
               0 \\ 0 \\ r\\ \alpha\left(1+r^{2}\right)
             \end{array} \right) ,
\end{eqnarray}
which does not satisfy the boundary condition at the $(2,1)$ edge;
it does not satisfy $H|\psi_{+}\rangle = 0$ on the two sites
designated by filled squares (blue) in Fig.~8(a).
Hence, we need to superpose $|\psi_{+}\rangle$ with other
zero-energy solutions of $H_{+}|\phi\rangle = 0$.
Indeed, we find two other solutions:
\begin{align}
 |\phi_{1}\rangle
  & = \sum_{i,j} \rho^{i}\left(-\rho^{2}\right)^{j} \, |i,j \rangle
      \left( \begin{array}{c}
               1 \\ 0 
             \end{array} \right) ,
     \\
 |\phi_{2}\rangle
  & = \sum_{i,j} \rho^{i}\left(-\rho^{2}\right)^{j} \, |i,j \rangle
      \left( \begin{array}{c}
               0 \\ 1
             \end{array} \right) ,
\end{align}
where $\rho$ satisfies $|\rho| < 1$ and $\Theta = 0$, that is,
\begin{align}
      \label{eq:Theta=0}
  \Theta = 1 + \alpha^{2} \left(r - \rho\right)\left(r - \rho^{-1}\right) = 0 .
\end{align}
Here, $\Theta = 0$ directly ensures $H_{+}|\phi_{1}\rangle = 0$ and
$H_{+}|\phi_{2}\rangle = 0$ as is apparent from the definition of $H_{+}$.
Equation~(\ref{eq:Theta=0}) with $|\rho| < 1$ yields
\begin{align}
    \label{eq:rho-def}
  \rho =  \frac{1}{2}
          \left( r+\frac{1}{r}+\frac{1}{\alpha^{2}r}
            -\sqrt{\left(r+\frac{1}{r}+\frac{1}{\alpha^{2}r}\right)^{2}-4}
          \right) .
\end{align}
The latter solution $|\phi_{2}\rangle$,
which is rewritten in the original $4 \times 4$ space as
\begin{align}
     \label{eq:phi}
 |\phi\rangle
    = \sum_{i,j} \rho^{i}\left(-\rho^{2}\right)^{j} \, |i,j \rangle
      \left( \begin{array}{c}
               0 \\ 0 \\ \alpha\left(r-\rho\right) \\ -1
             \end{array} \right) ,
\end{align}
is appropriate for the superposition because its amplitude is nonzero
on the third and fourth sites as in $|\psi_{+}\rangle$.
A general solution for the zero-energy edge state is
\begin{align}
    \label{eq:g-solution}
  |\tilde{\psi}\rangle = |\psi_{+}\rangle + \xi|\phi\rangle ,
\end{align}
where $|\psi_{+}\rangle$ and $|\phi\rangle$ do not satisfy
the original eigenvalue equation on the two sites
designated by filled squares (blue) in Fig.~8(a).
We determine $\xi$ such that $|\tilde{\psi}\rangle$ satisfies
$H|\tilde{\psi}\rangle = 0$ on the two sites, and find that
$\xi = \alpha(1-r\rho)$.
That is, the zero-energy edge state is expressed as
\begin{align}
       \label{eq:psi_{+}(2,1)}
  |\tilde{\psi}\rangle = |\psi_{+}\rangle + \alpha(1-r\rho)|\phi\rangle .
\end{align}
We can show that $|\tilde{\psi}\rangle$ vanishes
in the limit of $r \to 0$ except on the third and fourth sites
in each dual cell located at a $270^\circ$ corner.

One may think that the degrees of freedom of Eq.~(\ref{eq:g-solution}) are
insufficient for $|\tilde{\psi}\rangle$ satisfying the boundary condition.
Indeed, Eq.~(\ref{eq:g-solution}) contains only one free parameter $\xi$,
whereas the boundary condition requires
$H|\tilde{\psi}\rangle = 0$ on the two sites.
This seeming discrepancy is ascribed to the fact that
we adopt Eq.~(\ref{eq:condition2}) from the outset.
A more precise derivation of $|\tilde{\psi}\rangle$ is presented
in Appendix.
%%%%%%%%%%%%%%%%%%
\begin{figure}[btp]
\begin{center}
\includegraphics[height=7.0cm]{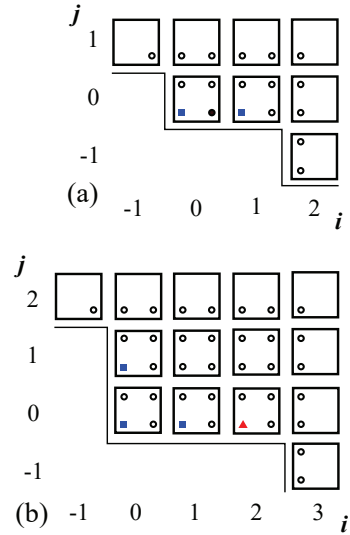}
\end{center}
\caption{
(Color online) Partial enlarged views of (a) Fig.~2(b) and (b) Fig.~2(c).
}
\end{figure}
%%%%%%%%%%%%%%%%%%

Let us consider the bulk--boundary correspondence in this case.
The only difference from the $(1,1)$ edge case is that
$\Theta$ in $H_{\pm}$ depends on $k$ as
\begin{align}
  \Theta = 1 + \alpha^{2}\left|r-e^{ika}\right|^{2} .
\end{align}
Because $\Theta > 0$ for $k \in [0,\frac{2\pi}{a})$,
$\Theta$ does not affect a gap closing and therefore
plays no role in the argument of bulk--boundary correspondence.
Thus, we can show that the bulk--boundary correspondence holds
in the $(2,1)$ edge case by repeating the argument given in Sect.~5
with $w_{\pm}$ defined in Eqs.~(\ref{eq:def_w-}) and (\ref{eq:def_w+}).

Let us turn to the $(3,2)$ edge as shown in Fig.~2(c).
Numerical results (data not shown in Sect.~3) show that the edge states
become gapless under the condition
\begin{align}
   \label{eq:condition3}
 \left(\frac{\gamma_{x}}{\lambda_{x}}\right)^{3}
 = \left(\frac{\gamma_{y}}{\lambda_{y}}\right)^{2}= r^{6}
\end{align}
with $0 \le r < 1$, and that two edge states at zero energy change
their sign under the lattice translation of $(\pm 3a, \mp 2a)$.
To describe the zero-energy edge states, we impose Eq.~(\ref{eq:condition3})
on $H(k_{x},k_{y})$ and replace $k_{x}$ with $2k+k_{\pi}$
and $k_{y}$ with $3k$, resulting in
\begin{align}
   Q
  = \left[ 
      \begin{array}{cc}
        r^{2}-e^{i2ka} & \alpha \left(r^{3}+e^{i3ka} \right) \\
         -\alpha \left(r^{3}+e^{-i3ka} \right) & r^{2}-e^{-i2ka}
      \end{array}
     \right] .
\end{align}
This is rewritten as
\begin{align}
   Q
  = \left[ 
       \begin{array}{cc}
         \Sigma \eta & \Gamma \eta \\
         -\Gamma^{*}\eta^{*} & \Sigma^{*}\eta^{*}
       \end{array}
     \right] ,
\end{align}
where
\begin{align}
  \eta & = r + e^{ika} , \\
  \Sigma & = r - e^{ika} , \\
  \Gamma & = \alpha \left(r^{2} - re^{ika} + e^{i2ka} \right) .
\end{align}

The one-dimensional Hamiltonians $H_{+}$ and $H_{-}$ given in
Eqs.~(\ref{eq:H+}) and (\ref{eq:H-}), respectively,
are also applicable to this case.
Two zero-energy edge states are obtained by solving
$H_{+}|\psi_{+}\rangle = 0$ and $H_{-}|\psi_{-}\rangle = 0$
after the replacement of
$e^{ika}$ with $\rho$ and $e^{-ika}$ with $\rho^{-1}$.
From $H_{-}|\psi_{-}\rangle = 0$, we find a solution with $\rho = -r$
and the eigenvector $^{t}(1, 0)$ corresponding to $v_{-}^{(1)}$.
This is expressed as
\begin{align}
    \label{eq:psi_{-}(3,2)}
 |\psi_{-}\rangle
    = \sum_{i,j} (-r^{2})^{i}(-r^{3})^{j} \, |i,j \rangle
      \left( \begin{array}{c}
               0 \\ 1 \\ 0\\ 0
             \end{array} \right) ,
\end{align}
which satisfies the boundary condition at the $(3,2)$ edge.
We observe that $|\psi_{-}\rangle$ vanishes in the limit of $r \to 0$
except on the second site in each dual cell located at a $90^\circ$ corner.

To find another zero-energy edge state originating from $270^\circ$ corners,
we solve $H_{+}|\psi_{+}\rangle = 0$ and find a solution with
$\rho = -r$ and the eigenvector $^{t}(0, 1)$ corresponding to $v_{+}^{(2)}$.
This is expressed as
\begin{align}
 |\psi_{+}\rangle
    = \sum_{i,j} (-r^{2})^{i}\left(-r^{3}\right)^{j} \, |i,j \rangle
      \left( \begin{array}{c}
               0 \\ 0 \\ r+\frac{1}{r} \\
                    \alpha \left(r^{2}+1 +\frac{1}{r^{2}}\right)
             \end{array} \right) .
\end{align}
Because this does not satisfy $H|\psi_{+}\rangle = 0$ on the four sites
designated by filled squares (blue) and a filled triangle (red) in Fig.~8(b),
we need to superpose $|\psi_{+}\rangle$ with other
zero-energy solutions of $H_{+}|\phi\rangle = 0$.
By making a procedure similar to the derivation of Eq.~(\ref{eq:phi}),
we find two other solutions:
\begin{align}
      \label{eq:phi_(3,2)}
 |\phi_{\pm}\rangle
  & = \sum_{i,j} \left(-\rho_{\pm}^{2}\right)^{i}
                 \left(\rho^{3}_{\pm}\right)^{j}
           \nonumber \\
  & \hspace{5mm} \times
      |i,j \rangle
      \left( \begin{array}{c}
               0 \\ 0 \\ \alpha \left(r^{2}-r\rho_{\pm}+\rho_{\pm}^{2}\right)
                 \\ \rho_{\pm}-r
             \end{array} \right) ,
\end{align}
where $\rho_{+}$ and $\rho_{-}$ satisfy $|\rho_{\pm}| < 1$ and
\begin{align}
          \label{eq:Theta=0_32}
  \Theta =
  & \left(r-\rho_{\pm}\right)\left(r-\rho_{\pm}^{-1}\right)
        \nonumber \\
  & + \alpha^{2} \left(r^{2} - r\rho_{\pm} + \rho_{\pm}^{2}\right)
                 \left(r^{2} - r\rho_{\pm}^{-1} + \rho_{\pm}^{-2}\right) = 0 .
\end{align}
Equation~(\ref{eq:Theta=0_32}) with $|\rho_{\pm}| < 1$ yields
\begin{align}
     \label{eq:def^rho_pm}
  \rho_{\pm} = \frac{1}{2}\left(g_{\pm} + \sigma\sqrt{g_{\pm}^{2}-4}\right) ,
\end{align}
where $\sigma = +1$ or $-1$,
\begin{align}
    \label{eq:g-def}
  g_{\pm} =
  & \frac{1}{2\alpha^{2}r}
    \biggl\{1+\alpha^{2}\left(1+r^{2}\right)
        \nonumber \\
  & \hspace{-8mm}
       \pm \sqrt{\left(1+\alpha^{2}\left(1+r^{2}\right)\right)^{2}
                -4\alpha^{2}\left(1+r^{2}
                            +\alpha^{2}\left(1-r^{2}+r^{4}\right)\right)}
    \biggr\} .
\end{align}
In Eq.~(\ref{eq:def^rho_pm}), $\sigma$ is determined
such that $|\rho_{\pm}| < 1$.
A general solution for the zero-energy edge state is expressed as
\begin{align}
  |\tilde{\psi}\rangle = |\psi_{+}\rangle
   + \xi_{+}|\phi_{+}\rangle + \xi_{-}|\phi_{-}\rangle
   + \zeta|\varphi\rangle ,
\end{align}
where $|\varphi\rangle$ is defined by
\begin{align}
  |\varphi\rangle = \sum_{i=-\infty}^{+\infty}
                    (-1)^{i}|3i,-2i \rangle
                    \left( \begin{array}{c}
                             0 \\ 0 \\ \alpha r \\ -1
                           \end{array} \right) ,
\end{align}
having a nonzero amplitude only in each unit cell
located at a $90^\circ$ corner.~\cite{comment2}
Note that $H|\psi_{+}\rangle = 0$ and $H|\phi_{\pm}\rangle = 0$ do not hold
on the four sites designated by filled squares (blue)
and a filled triangle (red) in Fig.~8(b),
and $H|\varphi\rangle = 0$ does not hold
on the three sites designated by filled squares (blue).
Hence, $|\tilde{\psi}\rangle$ satisfies
$H|\tilde{\psi}\rangle = 0$ everywhere in the system
except on these four sites.
We determine $\xi_{+}$, $\xi_{-}$, and $\zeta$ such that
$|\tilde{\psi}\rangle$ satisfies $H|\tilde{\psi}\rangle = 0$
on the four sites, and we find
\begin{align}
   \xi_{\pm}
   & = \mp \frac{\alpha \left(1-r\rho_{\pm}+r^{2}\rho_{\pm}^{2}\right)}
                {r\left(r-\rho_{\pm}\right)\Pi}
       \Bigl\{  r\left(1+r^{2}\right)\rho_{\mp}
              + \left(1-r\rho_{\mp}\right)\rho_{\mp}^{2}
             \nonumber \\
   & \hspace{10mm}
              - \alpha^{2}\left(1+r^{2}\right)\left(1-r\rho_{\mp}\right)
                \left(r^{2}-\rho_{\mp}^{2}\right)
       \Bigr\}, \\
   \zeta
   & = \frac{\alpha \left(\rho_{-}-\rho_{+}\right)}{r^{2}\Pi}
       \Bigl\{  r\left(1-r\rho_{-}\right)\left(1-r\rho_{+}\right)
                \left(\rho_{-}+\rho_{+}\right)
              \nonumber \\
   & \hspace{0mm}
              + \left(1+r^{2}\right)
                \left[r^{2}
                      +\rho_{-}\rho_{+}
                       \left(1-r\left(\rho_{-}+\rho_{+}\right)\right)
                 \right]
       \Bigr\} ,
\end{align}
where
\begin{align}
   \Pi = & \rho_{-}\rho_{+}\left(\rho_{-}-\rho_{+}\right)
           \left(1-r\left(\rho_{-}+\rho_{+}\right)\right)
              \nonumber \\
         & \hspace{0mm}
       - \alpha^{2}r\left(1-r\rho_{-}\right)\left(1-r\rho_{+}\right)
         \left(\rho_{-}^{2}-\rho_{+}^{2}\right) .
\end{align}
We can show that
the resulting zero-energy edge state vanishes in the limit of $r \to 0$
except on the third and fourth sites in each dual cell
located at a $270^\circ$ corner.

As in the $(2,1)$ edge case,
\begin{align}
  \Theta = \left|r-e^{ika}\right|^{2}
           + \alpha^{2}\left|r^{2}-re^{ika}+e^{i2ka}\right|^{2}
\end{align}
plays no role in the argument of bulk--boundary correspondence.
We can show that the bulk--boundary correspondence holds
in the $(3,2)$ edge case by repeating the argument given in Sect.~5.

\section{Summary and Discussion}

A second-order topological insulator on a square lattice
hosts two families of one-dimensional edge states
when its edge is in a staircase form.
The spectrum of edge states tends to have a gap
owing to the hybridization between the two families.
However, under the condition given in Sect.~3, the spectrum becomes gapless
with a pair of zero-energy edge states at a gapless point.
This suggests that a certain symmetry
prohibits the hybridization and protects the pair of zero-energy edge states.
We identify the symmetry as a generalized mirror-rotation symmetry
and show that the bulk--boundary correspondence holds
in this system with a staircase edge.

Here, we briefly show that the results of analysis in this paper are valid
even when $N_{x}$ and $N_{y}$ become very large.
A crucial point is that we can determine a wavefunction of
a zero-energy edge state originating from $90^\circ$ corners
for arbitrary $N_{x}$ and $N_{y}$.
Our argument indicates that a pair of zero-energy edge states appear when
\begin{eqnarray}
    \left(\frac{\gamma_{x}}{\lambda_{x}}\right)^{N_{x}}
  = \left(\frac{\gamma_{y}}{\lambda_{y}}\right)^{N_{y}}= r^{N_{x}N_{y}}
\end{eqnarray}
with $r< 1$.
Using the procedure described in Sects~4 and 6,
we determine a wavefunction of a zero-energy edge state
originating from $90^\circ$ corners as
\begin{align}
 |\psi_{-}\rangle
    = \sum_{i,j} (-r^{N_{y}})^{i}(-r^{N_{x}})^{j} \, |i,j \rangle
      \left( \begin{array}{c}
               0 \\ 1 \\ 0\\ 0
             \end{array} \right) ,
\end{align}
which satisfies the boundary condition at the ($N_{x}$, $N_{y}$) edge.
We observe that $|\psi_{-}\rangle$ vanishes in the limit of $r \to 0$
except on the second site in each dual cell located at a $90^\circ$ corner.
Thus, by repeating the argument given in Sect. 5, we can show that
the bulk--boundary correspondence holds in the subspace described by $H_{-}$.
That is, a zero-energy edge state originating from $90^\circ$ corners
appears even when $N_{x}$ and $N_{y}$ become very large,
indicating that the hybridization between the two families of edge states
is prohibited by a generalized mirror-rotation symmetry.
From this argument, we expect the presence of a zero-energy edge state
originating from $270^\circ$ corners,
although its wavefunction is not easy to determine.

\section*{Acknowledgments}

This work was supported by JSPS KAKENHI
(Grant Numbers JP21K03405, JP17H06461, and JP19K14545),
JST CREST (Grant Number JPMJCR18T1),
and JST PRESTO (Grant Number JPMJPR19L7).

\appendix

\section{}

In this Appendix, we derive $|\tilde{\psi}\rangle$
given in Eq.~(\ref{eq:psi_{+}(2,1)}) on the basis of Eq.~(\ref{eq:Q2-init})
without assuming Eq.~(\ref{eq:condition2}) from the outset.
To do this, we set $r_{x}=\gamma_{x}/\lambda_{x}$ and
$r_{y}=\gamma_{y}/\lambda_{y}$
and treat $r_{x}$ and $r_{y}$ as independent parameters.

A fundamental solution for the zero-energy edge state
originating from $270^{\circ}$ corners is written as
\begin{align}
  |\Psi\rangle = \sum_{i,j} \rho^{i}(-\rho^{2})^{j}
                 |i,j \rangle
                 \left( \begin{array}{c}
                          0 \\ 0 \\ c \\ d
                        \end{array}
                 \right) ,
\end{align}
where $c$, $d$, and $\rho$ are determined by
\begin{align}
    \left[
      \begin{array}{cc}
        r_{x}+\rho & \alpha \left(r_{y}-\rho^{2}\right) \\
        -\alpha \left(r_{y}-\rho^{-2}\right) & r_{x}+\rho^{-1} \\
      \end{array}
    \right]
    \left( \begin{array}{c}
              c \\ d \\
           \end{array}
    \right)
    = 0 .
\end{align}
By solving this equation, we find two solutions
$|\Psi_{+}\rangle$ and $|\Psi_{-}\rangle$ for which
\begin{align}
  d_{\pm}
    = \frac{\alpha\left(r_{y}-\rho_{\pm}^{-2}\right)}{r_{x}+\rho_{\pm}^{-1}}
      c_{\pm}
\end{align}
and
\begin{align}
    \label{eq:rho-Appendix}
  \rho_{\pm} = \frac{1}{2}\left(h_{\pm} + \sigma\sqrt{h_{\pm}^{2}-4}\right) ,
\end{align}
where $\sigma = +1$ or $-1$,
\begin{align}
  h_{\pm} =
  & \frac{1}{2\alpha^{2}r_{y}}
    \biggl\{ r_{x}
        \nonumber \\
  & \hspace{-2mm}
       \pm \sqrt{r_{x}^{2}
                +4\alpha^{2}r_{y}
                 \left(1+r_{x}^{2}+\alpha^{2}\left(1+r_{y}\right)^{2}\right)}
    \biggr\} .
\end{align}
In Eq.~(\ref{eq:rho-Appendix}),
$\sigma$ is determined such that $|\rho_{\pm}| < 1$.
A general solution is written as
\begin{align}
  |\tilde{\Psi}\rangle = |\Psi_{-}\rangle + \xi|\Psi_{+}\rangle .
\end{align}
Imposing the boundary condition, which is equivalent to requiring
$H|\tilde{\Psi}\rangle = 0$ on every site on the $(2,1)$ edge,
we find that the boundary condition is satisfied only under the condition
of Eq.~(\ref{eq:condition2}) with
\begin{align}
  \xi = - \frac{c_{-}\rho_{+}}{c_{+}\rho_{-}} .
\end{align}
Applying Eq.~(\ref{eq:condition2}) (i.e., $r_{x}^{2}=r_{y}=r^{2}$)
to Eq.~(\ref{eq:rho-Appendix}),
we can show that $\rho_{-}=-r$ and $\rho_{+}$ is equivalent to
$\rho$ given in Eq.~(\ref{eq:rho-def}).
Thus, we finally arrive at the zero-energy edge state
$|\tilde{\psi}\rangle$ given in Eq.~(\ref{eq:psi_{+}(2,1)}).


\begin{thebibliography}{99}
\bibitem{comment1} Here, $\Sigma$ and $\Gamma$ are introduced
for later convenience despite the fact that they are rather redundant
in this case.
\bibitem{comment2} Note that $|\varphi\rangle$ is equivalent to
$|\phi_{\pm}\rangle$ given in Eq.~(\ref{eq:phi_(3,2)})
in the limit of $\rho_{\pm} \to 0$, indicating that it is included
in the subspace governed by $H_{+}$.
\end{thebibliography}
\end{document}